\documentclass[aip,pof,sd,amsmath,amssymb,preprint]{revtex4}
\usepackage{multirow}
\usepackage{psfrag}
\usepackage{color,epsfig}
\usepackage{amssymb}
\usepackage{textcomp} 
\usepackage{setspace}
\usepackage{mathrsfs}
\usepackage{dcolumn}
\usepackage{bm}%
\usepackage{natbib}
\usepackage{setspace}
\newcommand{\pdif}[2]{\frac{\partial #1}{\partial #2}}

\newcommand{\oline}[1]{ \overline{#1}}

\newcommand{\Rho}[1]{\bar{\rho}}

\newcommand{\etal}{\emph{et al.}}

\newcommand{\fav}[1]{\widetilde{#1}}
\definecolor{Olivegreen}{rgb}{0.0,.5,0}

\begin{document}

\title[Analysis of higher-order statistics in reacting turbulent wall-jets]
{Statistical analysis of the velocity and scalar fields in reacting turbulent wall-jets \footnote{Version accepted for publication (postprint) on Physics of
 Fluids {\bf 27} 025102 (2015)}}

\author{Z. Pouransari}
\email{zeinab@mech.kth.se.}
\affiliation{ 
Linn\'e FLOW Centre, KTH Mechanics,
SE-100 44 Stockholm, Sweden
}%
\author{L. Biferale}%
\affiliation{ 
Department of Physics and INFN, University of Rome "Tor Vergata", Italy
}%
\author{A. V. Johansson}
\affiliation{ 
Linn\'e FLOW Centre, KTH Mechanics,
SE-100 44 Stockholm, Sweden%
}%
\begin{abstract}
The concept of local isotropy in a chemically reacting turbulent wall-jet 
flow is addressed using direct numerical simulation (DNS) data.
Different DNS databases with isothermal and exothermic reactions are examined.
The chemical reaction and heat release effects on the turbulent velocity,
passive scalar and reactive species fields are studied
using their probability density functions (PDF) and higher order moments
for velocities and scalar fields, as well as their gradients. 
With the aid of the anisotropy invariant maps for the Reynolds stress tensor
the heat release effects on the anisotropy 
level at different wall-normal locations are evaluated and
found to be most accentuated in the near-wall region.
It is observed that the small-scale anisotropies are
persistent both in the near-wall region and inside the jet flame.
Two exothermic cases with different Damk\"ohler number are examined
and the comparison revealed that the Damk\"ohler number effects are
most dominant in the near-wall region, where the wall cooling effects are influential. 
In addition, with the aid of PDFs conditioned on the mixture fraction, the
significance of the reactive scalar characteristics in the reaction zone is illustrated.
We argue that the combined effects of strong intermittency and strong persistency of
anisotropy at the small scales in the entire domain can affect
mixing and ultimately the combustion characteristics of the reacting flow.
\end{abstract}
\maketitle
\section{Introduction}
The concept of local isotropy is important for both fundamental and applied problems in turbulence~\citep{shen_war,biferale2005,kurien_sreenivasan_etal,Cambon2009}.
On the one hand, obviously no real flow can be exactly isotropic at all scales,
due to the external forcings and boundary conditions that typically break such symmetry.
On the other hand, theoretical ideas and experimental
observations support the concept of 'return to local isotropy',
where by local it is meant 'far enough from the boundaries' and also for 'small enough' scales.
The rationale is that any anisotropic effects introduced by the external forcing, 
the mean flow and/or walls, is overwhelmed by the turbulent
fluctuations at sufficiently large Reynolds numbers and
in fluid locations where neither the forcing nor the boundaries play a major role (if any).
Therefore, all flows in nature have some degree of anisotropy, but also all
flows in nature should return to some quasi-isotropic statistical state.
Moreover, the return to isotropy should be faster than the decay rate of turbulence.
Even though there are many confirmations of this general phenomenology
(see e.g.\citet{biferale2005} for a review) there are still some crucial open questions.

Concerning scalar and temperature gradient fluctuations, 
the available data and discussions are even more diverse
and complicated than for the turbulent velocity.
Already, the behavior of the passive scalar statistics is
non-trivial and still a major field of research\citep{falkovich2001}.
The complex aspects of the scalar characteristics apparently arise
from the mixing process rather than from the turbulent velocity
field\cite{shraiman} and the scalar field is mainly decoupled from the velocity field itself.
The proper way to assess the relative importance of isotropic {\it vs} anisotropic fluctuations in
turbulent flows, including active or passive scalar advection,
is given through the direct measurements of the moments of the 
increments of the fields for different distances, i.e. the so-called structure functions\cite{frisch}.
Indeed, the behavior of the higher order moments of small-scale fluctuations in turbulent flows
has been the subject of both numerical and experimental studies for decade\citep{gotoh_kaneda,biferale2010,benzi,sreenivasan97,meneveau_marusic,Cambon2013,Gylfason_warhaft,schumacher2003}.
It is crucial to realize that not all possible structure functions are sensitive to anisotropy in the same way. 
For instance, it is well known that odd moments of scalar fields are always vanishing for perfectly isotropic statistics,
while for velocity increments only those built in terms of the transverse components must vanish\citep{biferale2001,kurien_sreenivasan_etal,kurien_sreenivasan,lvov_procaccia,shen_war,biferale_toschi}.
Therefore, when a stable statistical signal is measured for an odd increment
of a scalar field or a transverse velocity component in a turbulent flow,
it is a direct signature of the presence of some kind of anisotropy.
On the other hand, the even moments are less sensitive, being always
different from zero already in a fully isotropic ensemble.
From a theoretical point of view, there exists a very clear and systematic way
to disentangle isotropic and anisotropic fluctuations, and to
disentangle different kinds of anisotropic fluctuations among themselves.
This is based on the idea to decompose any observable in a suitable
set of eigenfunctions of the group of rotation\citep{biferale2005,arad98}.
As a matter of fact, it is very difficult to apply this decomposition on experimental data,
because of the need to control the whole field on a spherical volume.
This is why systematic studies of the anisotropic properties of turbulent flows are rare.
In addition, in order to better understand the whole
probability density function (PDF) of anisotropic fluctuations,
it is mandatory to also measure high order moments, a task
that can only be accomplished with large statistical samples.
This is why the combined assessment of importance of anisotropic
and rare non-Gaussian fluctuations (intermittency) in turbulent flows and
scalar fields advected by turbulent flows is still considered a state-of-the-art theoretical and experimental problem\citep{antonia73,bailey2013,talamelli,dewit,Gylfason_warhaft,antonia81,Ferchichi_Tavoularis,antonov}.
The situation is even more complex when the scalar field is coupled
back to the advecting velocity, becoming an active component in the
dynamics\cite{celani_cencini}, as for the case of the turbulent Rayleigh-Benard convection,
where the temperature preferentially couples with the velocity fluctuations in the
vertical direction only, introducing a strong large scale anisotropy due to the buoyancy effects.

The reacting flows in general and combustion applications in particular are
among the turbulent flows, where large and small scales of different scalar fields
are present and coupled back to the advecting flow field.
Many researchers have studied the characteristics of
passive and active scalars embedded in turbulent reacting flows to detect
possible universal/non-universal behaviors\citep{pet,bilger2005}.
However, studies on the effects of the combustion heat release on the
small scales of turbulence are limited and need to be further explored.
For instance, Tong, Warhaft and coworkers showed that the
PDF of the scalar derivative in the direction of the mean gradient is strongly skewed.
They observed that the skewness of the scalar derivative
in the direction of the mean scalar gradient is nonzero,
which implies the existence of the ramp-cliff-like structure in the
scalar signal\citep{Tong_94,Tong_2007,Tong_2009,Tong_2011,Tong_2013}.
They also showed that the ramp-cliff structure in the mixture fraction
field has a strong influence on the reaction zones.
Tong, Barlow and coworkers\cite{Tong_2009} properly asserted that
the understanding of turbulent mixing is largely based on the
Kolmogorov turbulence theory.
They discussed that there are certain deviations from the universal
distribution of the scalar at the subgrid-scale (SGS) level.
They argued that in both reacting and non-reacting flows, depending on
the SGS scalar variance and mean value, the SGS scalar is expected to
show either a nearly Gaussian or a non-Gaussian behavior.
They categorized the scalar structure at the SGS
level to be a ramp-cliff structure\cite{Tong_2011}.
Their finding regarding the SGS scalar properties has many
phenomenological parallels with the results of the present
study concerning the small-scale characteristics.

In the present work, we analyze the case of an active scalar field in a
reacting turbulent wall-jet, which represents a complex canonical flow.
The scalar reacts and affects the flow and its influence
depends on the position, due to the spatial evolution of the jet and
boundary layer in the streamwise direction
and the chemical reaction that occurs between the species.
We investigate the heat release effects
on the turbulent flow, with an emphasis on the 
anisotropic and/or intermittent aspects.
Indeed, the heat-release effects on
the anisotropic fluctuations of the wall-jet flow and its 
impacts on the scalar fields are not yet well understood.
We study this by looking at both the large and small-scale statistics at different 
downstream locations and wall distances. 
\begin{figure}
\centering
\setlength{\unitlength}{\textwidth}
\includegraphics[width=12cm]{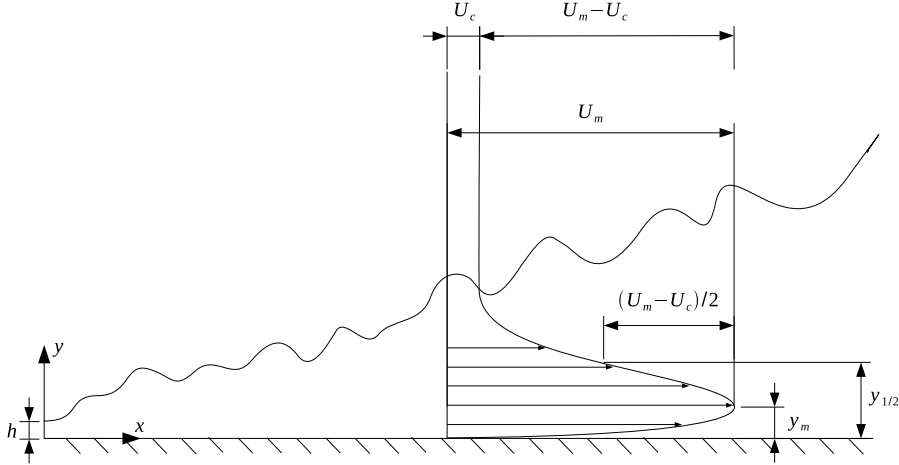}
\caption{Schematic of the plane turbulent wall-jet with coflow.
Here, $U_m$ is the local maximum streamwise velocity, $U_c$ is the coflow velocity,
$y_m$ is the distance from the wall at the position of maximum streamwise velocity
and $y_{1/2}$ is the half-height of the wall-jet in the outer region.} 
\label{fig:schemi}
\end{figure}

The turbulent wall-jet consists of a boundary layer and
a jet flow\citep{LaunderRodi1983,Dejoan_Lesh2005}, both of which are
among the most important and fundamental canonical turbulent flows, see figure~\ref{fig:schemi}.
Inclusion of the chemical reaction allows us to compare the statistics of the active and reactive scalar fields.
This is achieved by integrating, simultaneous to the reacting field,
also the evolution of another scalar, fully passive, without any
feedback on the flow.
In a previous study\cite{Pouransari2013},
we have examined and addressed the heat-release effects on the mixing scales
of the flow, which play a role on the large-scale features of the velocity and scalar fields.
In the present study, the DNS databases of two different reacting turbulent wall-jets are considered.
In one case\cite{Pouransari2011}, the chemical reaction is isothermal
while in the other\cite{Pouransari2013}, it is exothermic with a substantial amount of heat release.
Since, no heat release is involved in the isothermal case, the
reactive scalar is "passive-reactive", while for the
exothermic case, the reactive scalar is tagged as the "active-reactive" scalar.
Moreover, we also evolve a passive non-reacting scalar in the same flow configuration.
Thus, three different types of scalars are present, namely, passive,
passive-reactive and active-reactive scalars and the present investigation
addresses the differences in their behaviors.
The study of the small-scale characteristics of the turbulent
flow and the scalar field has interesting outcomes and implications
for turbulence modeling applications\cite{johansson2013,johansson2014}.

It is known that the chemical reaction leads to a strong non-Gaussian behavior at
large scales leading to different characteristics of the reactive
scalar from that of the passive scalar.
Chemical reaction enhances the intermittency level at
different locations, e.g. near the wall and at the
half-height of the jet, where the reaction rate is maximum.
As discussed in Pouransari \etal \citep{Pouransari2011}, the
PDFs of fuel species show long exponential tails.
Here, we further investigate whether the non-Gaussian
behavior caused by the chemical reaction
is persistent down to the very small scales. 
It is observed that, the persistency of anisotropy and
the related intense intermittency of both passive and active scalars
are greater than that of the velocity field in which it is embedded.
We measure the departure from local isotropy at small scales and
the degree of intermittency of the various scalars.
In order to gain a firm foundation for the understanding
of scalar intermittency and anisotropies, the behavior of higher order moments
for both velocity and scalar fields and their gradients are examined.

The paper is organized as follows.
The governing equations are given 
in section \ref{sec:gov} and the numerical method and flow 
configuration are explained in section \ref{sec:num}.
Mean flow statistics are briefly discussed in section \ref{sec:sim}.
In section \ref{sec:pdf}, PDFs for the velocity and scalar fields are discussed.
The PDFs of two different cases with different Damk\"ohler numbers are compared.
Moreover, the conditional PDFs are discussed with a focus on the flame zone.
The anisotropy invariant maps are used in section \ref{sec:anisot}
for an additional analysis of the anisotropy of the flow field.
Furthermore, discussions on the higher order moments of the velocity 
fields as well as different scalar fields are included in section \ref{sec:ho},
followed by an analysis of the gradient fields.
Conclusions are drawn in section \ref{sec:conc}.
\section{Governing equations for compressible reacting flow}\label{sec:gov}
The conservation equations of total mass, momentum and energy read
\begin{eqnarray}
 \label{eq:gov_eq}
 \pdif{\rho}{t} + \pdif{\rho u_{j}}{x_{j}} = 0 \\
 %
 \label{eq:mom}
 \pdif{\rho u_{i}}{t} + \pdif{\rho u_{i}u_{j}}{x_{j}} = 
 -\pdif{p}{x_{i}} + \pdif{\tau_{ij}}{x_j} \\
 %
 \label{eq:en}
 \pdif{\rho E}{t} + \pdif{\rho E u_{j}}{x_{j}} = \dot{\omega}_T
 -\pdif{q_i}{x_{i}} 
 +\pdif{(u_{i}(\tau_{ij}-p\delta_{ij}))}{x_{j}}.
\end{eqnarray}
Here $\rho$ is the total mass density, $u_i$ are the velocity 
components, $p$ is the pressure, $E=e+1/2\,u_i u_i$ is the 
total energy and $\dot{\omega}_T$ is the heat-release 
term due to the exothermic reaction (${\dot{\omega}_T}^{(i)}=0$ for the isothermal case).
The summation convention over repeated indices is used. 
The heat fluxes $q_i$ are approximated by 
Fourier's law $ q_i = - \lambda (\partial T/ \partial x_i) $, where $\lambda$ is 
the constant coefficient of thermal conductivity and $T$ is the temperature.
The viscous stress tensor is defined as
$\tau_{ij} = \mu\left( \partial u_i/ \partial x_j + \partial u_j/ \partial x_i
 \right) - \mu(2/3)(\partial u_k/ \partial x_k) \delta_{ij}$,
where $\mu$ is the dynamic viscosity and $\delta_{i,j}$ is the Kronecker delta.
The fluid is assumed to be calorically perfect and to obey the ideal gas law according to 
$e = c_v T$, $p = \rho R T$ and a specific heat ratio 
of $\gamma=c_p/c_v=1.4$ is used.
The viscosity is determined through the Sutherland's law
\begin{equation}
 \frac{\mu}{\mu_{jet}}=\left(\frac{T}{T_{jet}}\right)^{3/2}\frac{T_{jet}+S_0}{T+S_0},
\end{equation}
where $T$ is the local temperature, $T_{jet}$ is the jet center 
temperature at the inlet and $S_0=110.4K$ is a reference value.
Conservation of the species masses is governed by
\begin{eqnarray}
 \pdif{\rho \theta_o}{t} + \pdif{}{x_{j}}\left( \rho \theta_o u_{j}
 \right) 
 & = & \pdif{}{x_{j}} \left( \rho \mathcal{D} \pdif{\theta_o}{x_{j}}
 \right) - \dot{\omega}_o ,
 \\
 \pdif{\rho \theta_f}{t} + \pdif{}{x_{j}}\left( \rho \theta_f u_{j}
 \right) 
 & = &\pdif{}{x_{j}} \left( \rho \mathcal{D} \pdif{\theta_f}{x_{j}}
 \right) - \dot{\omega}_f ,
 \\
 \pdif{\rho \theta}{t} + \pdif{}{x_{j}}\left( \rho \theta u_{j}
 \right) 
 & = &\pdif{}{x_{j}} \left( \rho \mathcal{D} \pdif{\theta}{x_{j}}
 \right),
\end{eqnarray}
where $\theta_o$, $\theta_f$ and $\theta$ are the mass fractions 
of the oxidizer, fuel and passive conserved scalar, respectively and $\dot{\omega}_o$ and $\dot{\omega}_f$ are the 
reaction rates for the oxidizer and fuel species.
An equal diffusion coefficient $\mathcal{D}$ for scalars is
used to approximate the diffusive fluxes.\\
The chemical reaction between the oxidizer $O$ and fuel species $F$ is simplified as a single-step irreversible 
reaction that form the product $P$ as $\nu_O O + \nu_F F\longrightarrow \nu_P P$, 
whose rate is proportional to multiplication of the reactant concentrations
for the isothermal case and is expressed with the Arrhenius equation for the exothermic 
case, formulated as 
\begin{eqnarray}
 \label{eq:sr}
{\dot{\omega}_f}^{(i)}&=& -\, k_r \rho^2 \theta_o \theta_f,\\
\label{eq:arh}
{\dot{\omega}_f}^{(e)}&=& -\, k_r \, \rho^2 \,\theta_o \,\theta_f\, \exp(-Ze/T),
\end{eqnarray}
where ${\dot{\omega}_f}^{(i)}$ and ${\dot{\omega}_f}^{(e)}$ are the source terms in the fuel species
conservation equation for isothermal and exothermic cases, respectively, and $k_r$ is a
constant determining the Damk\"ohler number $Da=h/U_{jet}\, k_r \rho_{jet}$.
Note the difference in formulation of the reaction rate.
In the exothermic case, the chemical reaction is temperature
dependent with an exponential term as described in Eq.~(\ref{eq:arh}). A corresponding
source term as $\dot{\omega}_T$ appears in the energy equation, 
which is responsible for the combustion heat-release.
This will cause the major part of the thermal coupling between
the velocity and the scalar fields, apart from the minor compressibility effects.
In the isothermal case, the source term is independent
of the temperature, as expressed in Eq.~(\ref{eq:sr}) and
the energy equation is not affected by the chemical reaction.
The combustion heat-release term, $\dot{\omega}_T$ is related to 
the species reaction rates by
$\dot{\omega}_T = - \sum_{k=1}^{N} \Delta h_{f,k}^{0} \, \dot{\omega}_k$,
where $\Delta h_{f,k}^{0}$ is the formation enthalpy of the $k^{th}$-species,
for more details on governing equations see Pouransari \etal (2011) \cite{Pouransari2011}.

The half-height $y_{1/2}$ for the plane wall-jet is defined as the distance from the wall to the
position in the outer region, i.e. $y_{1/2} > y_{m}$,
where the mean velocity reaches half of the maximum excess value, i.e. $U(y_{1/2})=1/2(U_{m}-U_{c})$, 
where $U_m$ is the local maximum streamwise velocity, $y_m$ is its wall distance and $U_c$ is the coflow 
velocity, see figure~\ref{fig:schemi}.
With the spatial evolution of the turbulent jet flow, $y_m$ varies and thus $y_{1/2}$ will grow accordingly as is inherited in its definition.
The growth rate of the $y_{1/2}$ for the wall-jet flows is linear\cite{Pouransari2011} and is affected by the combustion heat release.
As described in Pouransari \etal (2013)\cite{Pouransari2013}, $y_{1/2}$ decreases moderately with introduction of heat release.
However, it can still serve as a valuable outer scaling parameter, when quantities in the outer region are discussed.
Therefore, throughout this study, $y_{1/2}$ is used for the outer scaling,
except where the combustion heat release affects the particular
parameter and the changes in $y_{1/2}$ with heat release need to be avoided.
Thus, in some of the illustrations, the jet inlet height $h$, which is a constant geometrical
parameter, is used as the normalization parameter for the two cases.

Two types of averaging have been employed in the paper.
The Reynolds averaged of $f$ is denoted by $\overline f$ using
$f=\overline f+f'$, where $f'$ is the fluctuation
about the mean value. Accordingly, when Favre averaging is 
used, $f$ is decomposed as $f=\widetilde f+f''$,
where $f''$ is the fluctuation about the density-weighted average 
and $\fav{f}=\overline{\rho f}/\overline \rho$ denotes the Favre-averaged mean value. 
\section{Numerical method, resolution and length scales}\label{sec:num}
\begin{figure}[tbh]
\begin{minipage}{\linewidth}
\centering
\includegraphics[width=\linewidth]{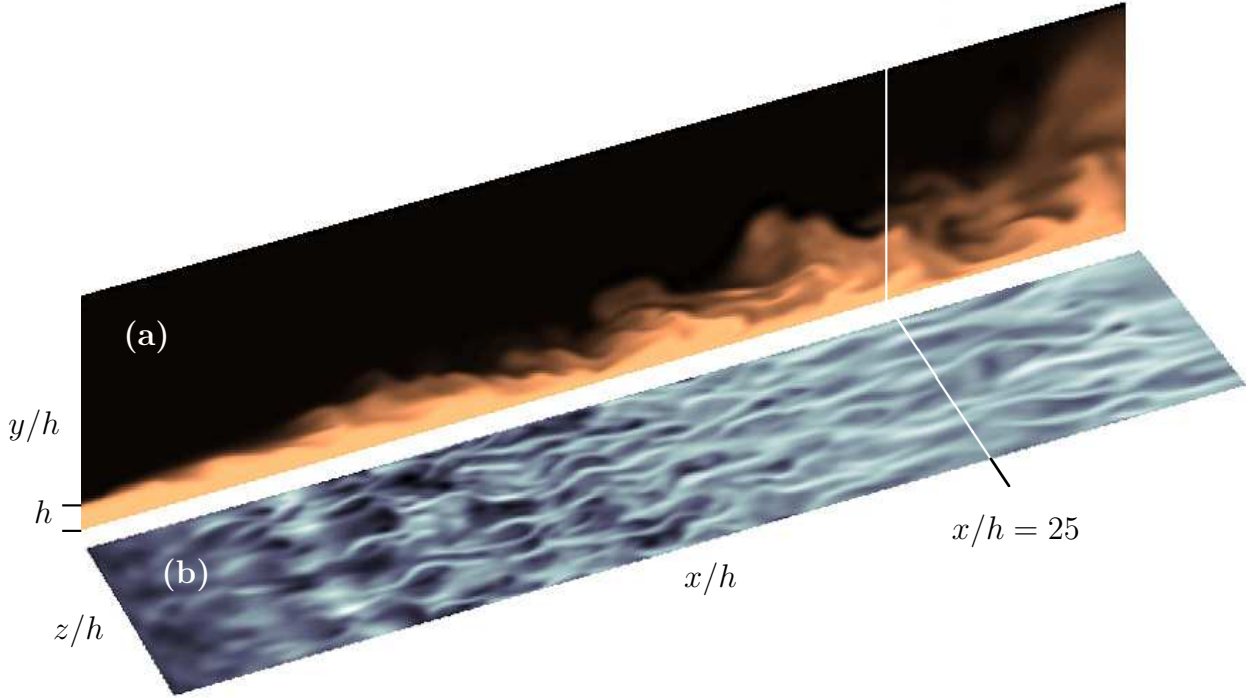}
\end{minipage}
\caption{
Instantaneous snapshots of (a) fuel concentration at an $x-y$ plane,
where a lighter color indicates a higher concentration
and (b) the streamwise velocity fluctuations at the $x-z$ plane at $y^+ \approx 8$.
Light and dark colors represent positive and negative velocity fluctuations, respectively.}
\label{fig:streak}
\end{figure}

\begin{table}
\begin{center}
\caption{Summary of the simulation cases, $L_i$ and $N_i$ are the domain size and 
grid points in the $i^{th}$-direction, $M$ is the Mach number and $Sc$ denotes the Schmidt number.}
\vspace{0.2cm}
\begin{tabular}{c|c|c|c|c|c|c|c|c}
\hline\hline
Case & reaction&$L_x\times L_y\times L_z$&$N_x\times N_y\times N_z$&Re&Da&Ze&M&Sc
\\\hline
I & isothermal\cite{Pouransari2011} &$35\times17\times7.2 $&$320\times192\times128$ &2000&3&0&0.5&0.72
\\\hline
II & exothermic\cite{Pouransari2013} &$35\times 17\times 7.2$&$320\times192\times128$&2000&1100&8&0.5&0.72
\\\hline
III & exothermic (lower Da)\cite{Pouransari2013} &$35\times 17\times 7.2$&$320\times192\times128$&2000&500&8&0.5&0.72
\\\hline\hline
\end{tabular} 
\end{center}
\label{tabl:cases}
\end{table}
A fully compressible Navier-Stokes solver \cite{Lele92,borsma99} is
employed for numerical simulation of the wall-jet flow.
The code uses a sixth-order compact 
finite difference scheme for spatial discretization and a third-order 
Runge-Kutta method for the temporal integration.
A schematic of the plane wall-jet is 
shown in figure~\ref{fig:schemi} and snapshots of the simulation results are
illustrated in figure~\ref{fig:streak}, where $h$ is the jet inlet height 
and $x$, $y$ and $z$ denote the streamwise, wall-normal and spanwise 
directions, respectively.

The computational domain is a rectangular box.
The extents of the geometry in terms of the inlet jet height $h$, and number
of grid points together with other simulation specifications are given in 
Table~I. Details for case~I can be found in Pouransari \etal~(2011)\citep{Pouransari2011} 
and for case~II and case~III in Pouransari \etal~(2013)\citep{Pouransari2013}.
Note that computation of the averaged statistics and PDFs are carried out using
both $N_{T}=1087$ snapshots and the data points in the spanwise homogenous direction with $N_z=128$ points.
Thus, in total $N_z\times N_T$ datasets are used for calculation of the averages, see Table 1.\\
The reactants enter the domain in a non-premixed manner. 
At the inlet of the computational domain, both fuel species
and passive scalar are injected through the jet stream, within the height $h$. 
The remaining part of the inlet boundary consists of a coflow
containing 50\% of oxidizer species, while the jet flow consists of 100\% of fuel species.
At the wall, the no-slip condition is fulfilled for the velocity and a 
no-flux condition, ${( \pdif{\theta}{y})}_{y=0}=0$, is applied for the 
scalars. Periodic boundary conditions are used in the spanwise 
direction. The ambient flow above the jet has a constant coflow 
velocity of $U_c = 0.10\,U_{jet}$, where $(jet)$ is used 
to denote properties at the inlet jet center and $U_{jet}$ is the jet inlet velocity.
At the top of the domain an inflow velocity of $0.026\,U_{jet}$ is used to 
account for the entrainment. To prevent the reflection and generation 
of waves, sponge zones are implemented at the inlet and outlet 
boundaries.

The Reynolds number is defined as $Re=U_{jet}h/\nu_{jet}$, where $\nu_{jet}$ is the jet inlet viscosity.
Mach and Schmidt numbers are defined as $M = U_{jet} /a$ and $Sc=\nu/\mathcal{D}$, respectively, 
where $\nu$ is the kinematic viscosity and $a$ is the speed of sound; 
and their numerical values are specified in Table~1. For the heat 
fluxes a constant Prandtl number $Pr = \mu c_p /\lambda = 0.72$ is 
used. The Schmidt number of the scalars is also constant and equal to 
the Prandtl number, i.e. $Sc = \mu/\rho D = 0.72$. 
For details concerning the resolutions, the 
boundary conditions and the numerical methodology,
see Pouransari \etal (2011,2013)\citep{Pouransari2011,Pouransari2013}.
In order to take into account the compressibility 
effects on the statistics, Favre averages are computed. However, most of the Favre averaged statistics 
do not significantly differ from their Reynolds-averaged values.

In the streamwise direction, grid stretching is used to give a higher 
resolution in the transition region. At the downstream position 
$x/h=25$, where most of the statistics presented in the following sections 
are taken, the streamwise and spanwise resolutions in wall units are $\Delta x^+\approx10$
and $\Delta z^+\approx6$, respectively. In the wall-normal direction, the grid is made finer near the 
wall, using a tangent hyperbolic distribution, to provide sufficient resolution for resolving the inner layer structures.

%
\begin{figure}[!t]
\centering
\includegraphics[width=\linewidth]{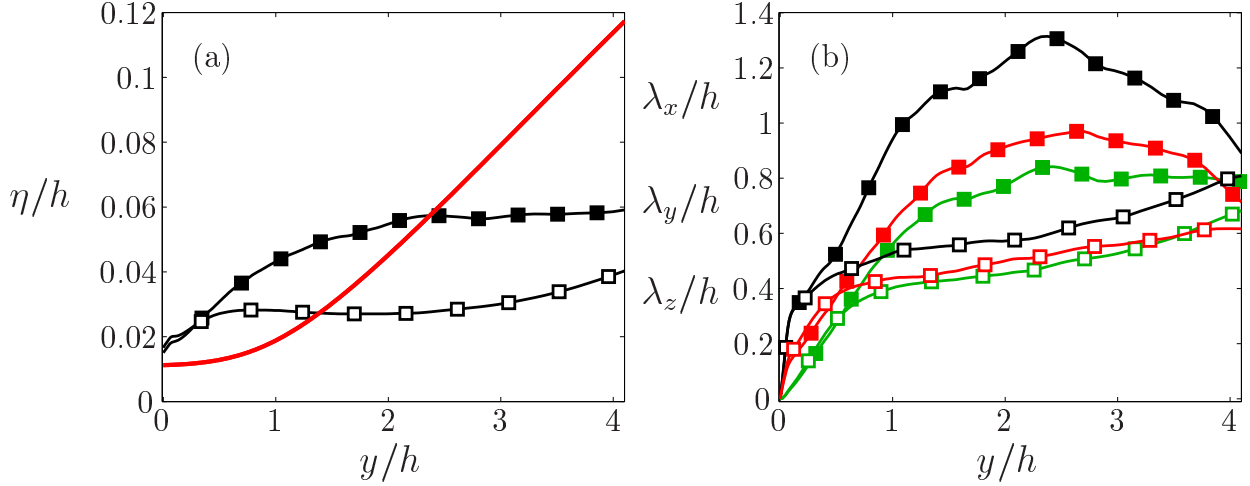}
\caption{\label{fig:kolm_tayl}
(a) Variation of the normalized Kolmogorov length scale $\eta/h$, h is the jet inlet height, with the wall distance. Black
unfilled squares, isothermal case I; black filled squares, exothermic case II; red solid curve, normalized wall-normal grid
spacing $\Delta y/h$. (b) Variation of the normalized Taylor micro-scale in the streamwise, wall-normal, and spanwise directions:
$\lambda_x/h$, $\lambda_y/h$, and $\lambda_z/h$, respectively, with the wall distance. Black unfilled squares, $\lambda_x/h$; red unfilled squares, $\lambda_y/h$; and
green unfilled squares, $\lambda_z/h$ for the isothermic case I. Black, red, and green colors are used for $\lambda_x/h$, $\lambda_y/h$, and $\lambda_z/h$.
Unfilled and filled symbols denote the isothermal case I and exothermic case II, respectively. Figures are at downstream
position x/h = 25.
}
\end{figure}
A Taylor micro scale in the $i^{th}$ direction can be defined as 
\begin{equation}
\label{eq:taylor}
\lambda_{i} =\sqrt{\frac{15 \nu}{\epsilon} u_{i,rms} ^ 2},\quad (i=1,2,3),
\end{equation}
where $\epsilon$ is the viscous dissipation rate. The Kolmogorov length scale is defined as
\begin{equation}
\label{eq:kolmo}
 \eta= \left(\frac{\nu ^3}{\epsilon}\right)^{1/4}.
\end{equation}
Figures~\ref{fig:kolm_tayl} (a) and (b) illustrate variation of these scales for the
current simulations in the wall-normal direction at $x/h=25$.
In this figure, a line with solid symbols is used for case II that includes the combustion heat release
and a line with empty symbols for the isothermal case I.
Note that these line styles are used throughout the paper unless otherwise stated.
Figure~\ref{fig:kolm_tayl} shows that $\lambda$ and $\eta$ become larger due to
the combustion heat release, both in the inner and outer regions of the flow,
which is consistent with the increase in viscosity and the other length scales of the
wall-jet flow, reported in Pouransari \etal (2013)\citep{Pouransari2013}.
Considering, the value of Taylor micro scale $\lambda_y$ for case~I at the
downstream position $x/h=25$, knowing that the half height of the jet is approximately
$y_{1/2}/h=2.5$\citep{Pouransari2013}, gives $\lambda_y/y_{1/2} \approx 1/5$.
This relatively large ratio is due to the low Reynolds number of the flow and
gets even larger for case~II with heat release. 
However, the friction Reynolds number of the flow is about $Re_{\tau}=230$,
which is comparable with that in typical low Reynolds number turbulent channel flow simulations\cite{kim}.
The local Reynolds number of the flow decreases with increasing viscosity due to the heat release.
This is the primary reason for both the micro scale and Kolmogorov scale to become larger
 in the exothermic case compared to the isothermal case\citep{pvj2014}.
Considering wall-normal variations of the streamwise and spanwise
Taylor micro scales $\lambda_x$ and $\lambda_z$, as
defined in Eq.~(\ref{eq:taylor}), a similar behavior is observed in all three directions of the flow.
From figures~\ref{fig:kolm_tayl}(a) and (b), we also observe that the small scales
of the flow are smaller at almost all wall-normal distances for the isothermal case.
However, further away from the wall, this reduction in the scale sizes is less accentuated.

\begin{figure}[!t]
\centering
\includegraphics[width=0.5\linewidth]{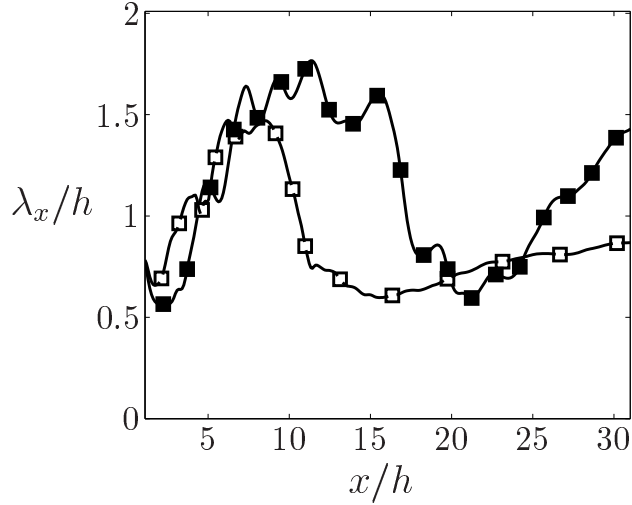}
\caption{\label{fig:kolm_tayl_x}
Streamwise variation of the normalized streamwise Taylor micro-scale $\lambda_x/h$ at the wall-normal position $y/h = 4$.
Here, $h$ is the jet inlet height. Black unfilled squares, isothermal case I and black filled squares, exothermic case II.
}
\end{figure}
An important point to notice is how the Kolmogorov scale compares to the
grid spacing in the vertical direction. The jet flame burns mostly below $y/h = 4$,
where most of the flame-turbulence interactions take place\citep{Pantano04}, see figure~\ref{fig:streak} for an overview of the field.
In this region, the grid spacing is either smaller than or in the order of the Kolmogorov scale
for both cases and the resolution is finer in case~II, which includes the heat release.
A similar trend is observed in the axial direction, see figure~\ref{fig:kolm_tayl_x},
where $\lambda_x$ increases with the heat release in general and the grid spacing is comparable
with the Kolmogorov scale. However, $\lambda_x$ reaches a maximum value in both cases I and II, close to the transition region at about $x/h =15$. Further downstream, $\lambda_x$ gradually decreases and at $x/h=25$, where it is well inside the self-similar region\cite{Pouransari2011},
the difference between the two cases is smaller and the heat release effects on the size of the small scales are minimum.
\section{Direct numerical simulation of turbulent wall-jet}\label{sec:sim}
To provide an overview of the turbulent wall-jet flow, visualizations of the
instantaneous passive scalar concentration for an $x-y$ plane
and streamwise velocity fluctuations for an $x-z$ plane at $y^+\approx8$ for
the isothermal reacting case I are shown in figures~\ref{fig:streak}(a) and (b), respectively.
The illustrations for the corresponding quantities of the exothermic case II are qualitatively similar and are not presented.
Note that figure~\ref{fig:streak}(a) shows a typical $x-y$ plane, which is
similar to other planes due to periodicity in the spanwise direction.
As observed in figure~\ref{fig:streak}(a), there exists a turbulent/non-turbulent interface
at the edge of the wall-jet flow, which is highly inhomogeneous.
The inhomogeneity at this interface is similar to that of the turbulent boundary layer flow; however,
this inhomogeneity exists similarly in both isothermal and exothermic cases.
Therefore, we can still compare the two cases in the entire domain and consider
the heat release effects on the characteristics of the flow at the interface region as well.

The near-wall streaky structures in figure~\ref{fig:streak}(b) indicate low-speed and
high-speed regions in the fluctuation field. The visualizations show that the jet is 
fully turbulent beyond $x/h\approx15$. Most of the statistics presented in this paper are taken at the
downstream position $x/h = 25$, where it is well inside the
fully developed region and also is far enough from the outlet boundary.

\begin{figure}[thb]
\centering
\includegraphics[width=\linewidth]{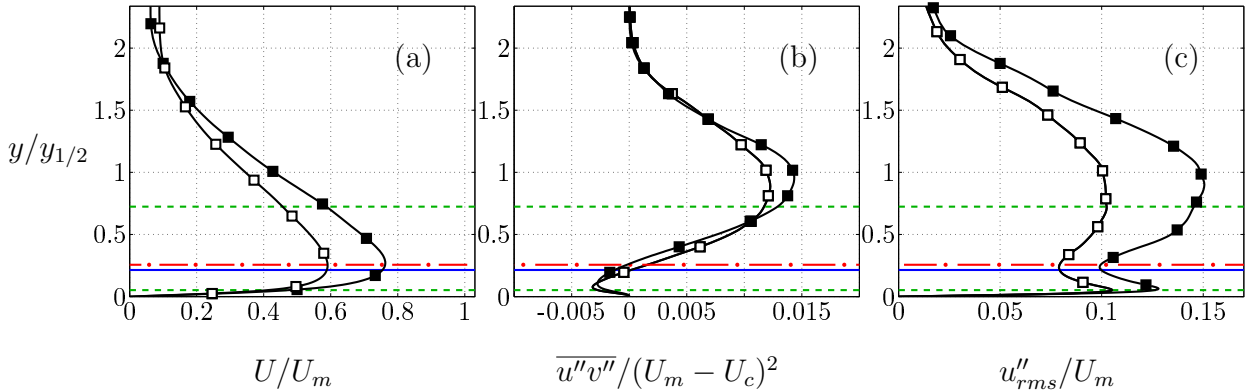}
\caption{
Cross-stream profiles of the Favre-averaged (a) mean streamwise velocity, (b) Reynolds shear stress, and (c)
streamwise velocity fluctuation intensity at $x/h = 25$ using outer scaling. Black unfilled squares, isothermal case I and black
filled squares, exothermic case II. The horizontal lines indicate particular wall-normal levels: red dotted-dashed line, the
position of the maximum mean streamwise velocity; blue solid line, the local zero shear stress position; and green dashed
lines, the positions of the streamwise fluctuation intensity maxima for the isothermal case I.
\label{fig:umax}
}
\end{figure}
Figures~\ref{fig:umax}(a-c) show the mean streamwise velocity,
Reynolds shear stress and the streamwise fluctuation intensity profiles, respectively, for the two
cases.
Some of the particularly interesting wall-normal positions for the isothermal case~I are indicated with horizontal lines.
Though the corresponding horizontal lines for the exothermic case~II are not shown in this figure, the discussion is valid for that case as well.
Note that, the position of the maximum streamwise velocity and the 
position of zero shear stress in the turbulent wall-jet flow do not exactly coincide.
This behavior has some similarities to what has been previously reported for jet flows.
We observe in figures~\ref{fig:umax}(a) and (b) that in the self-similar region ($x/h > 20$), the latter
lies below the former, and a region with negative mean production exists between them.
The zero crossing in the skewness of the streamwise velocity gradient fluctuations also lies
close to these positions, which will be discussed later in the discussion of figure~\ref{fig:skfl_du}(a).
\begin{figure}[thb]
\centering
\includegraphics[width=\linewidth]{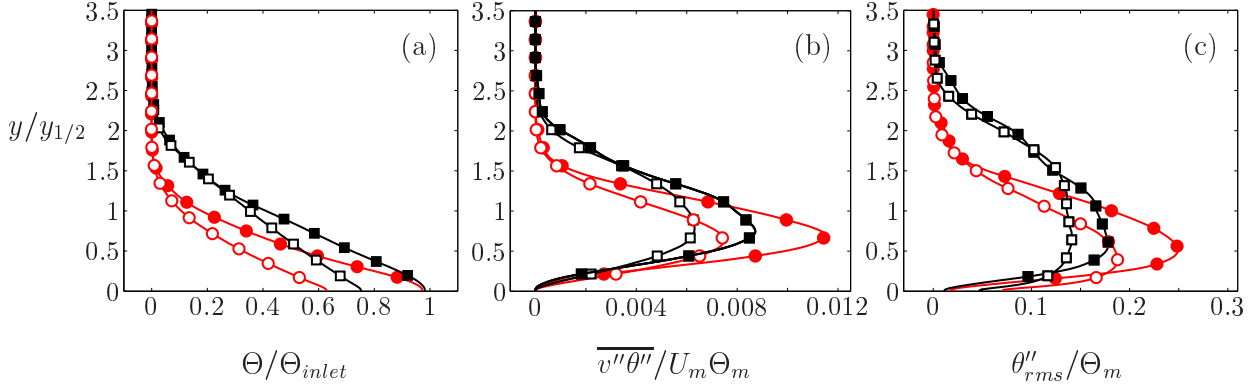}
\caption{
Cross-stream profiles of the Favre-averaged (a) mean, (b) wall-normal fluxes, and (c) fluctuation intensities of scalars
at downstream position $x/h = 25$ in outer units; black squares, passive scalar; red circles, fuel. Unfilled and filled symbols
denote the isothermal case I and exothermic case II, respectively. Normalization by the local maximum values $U_m$ and $\Theta_m$
of each case is used and $\Theta_{inlet}$ is the inlet scalar values.
\label{fig:ymean}
}
\end{figure}

Figures~\ref{fig:ymean}(a-c) show the cross-stream profiles of mean scalar concentrations, wall-normal flux and the fluctuation
intensities of both passive and reacting scalars for the two cases.
The filled and empty symbols are used for the exothermic and isothermal cases, respectively.
From these plots, it is deduced that both the chemical reaction itself and addition of the combustion heat release to the
reaction have significant effects on the statistics of the scalars.
That is to say, comparing the passive scalar statistics in the isothermal
reacting case I with that in the exothermic case II, i.e. considering
the heat release case, and then comparing the passive scalar with the reacting scalar within the
same simulation, i.e. considering the reaction effects, the changes due to both
effects are substantial.
Moreover, the influence of these two effects on the mean statistics seems to be of the same order.
\section{Probability density functions} \label{sec:pdf}
Probability density functions (PDFs) of the velocity components and different types of
scalars are examined at various wall-normal planes at downstream
position $x/h=25$ and for different observables.
We will mainly use the PDF of single point quantities to assess the large-scale fluctuations and
the PDF of gradients of the different fields to assess the small-scales properties.
In both cases, we will focus both on isotropic and anisotropic quantities.
The shape of the PDFs of the velocities and scalars and their derivatives
can assist us for the evaluation of these fields\cite{iti2010,iti2012}.
For instance, the long tails of the PDFs is a measure for the intermittency
and the number of local peaks indicates different modes of the reacting flow.
The presence of an exponential tails can be of particular
importance in practical applications for 
combustion modeling\cite{Warhaft,thmt,vervisch1995,wall2000}.
\begin{figure}[thb]
\centering
\includegraphics[width=\linewidth]{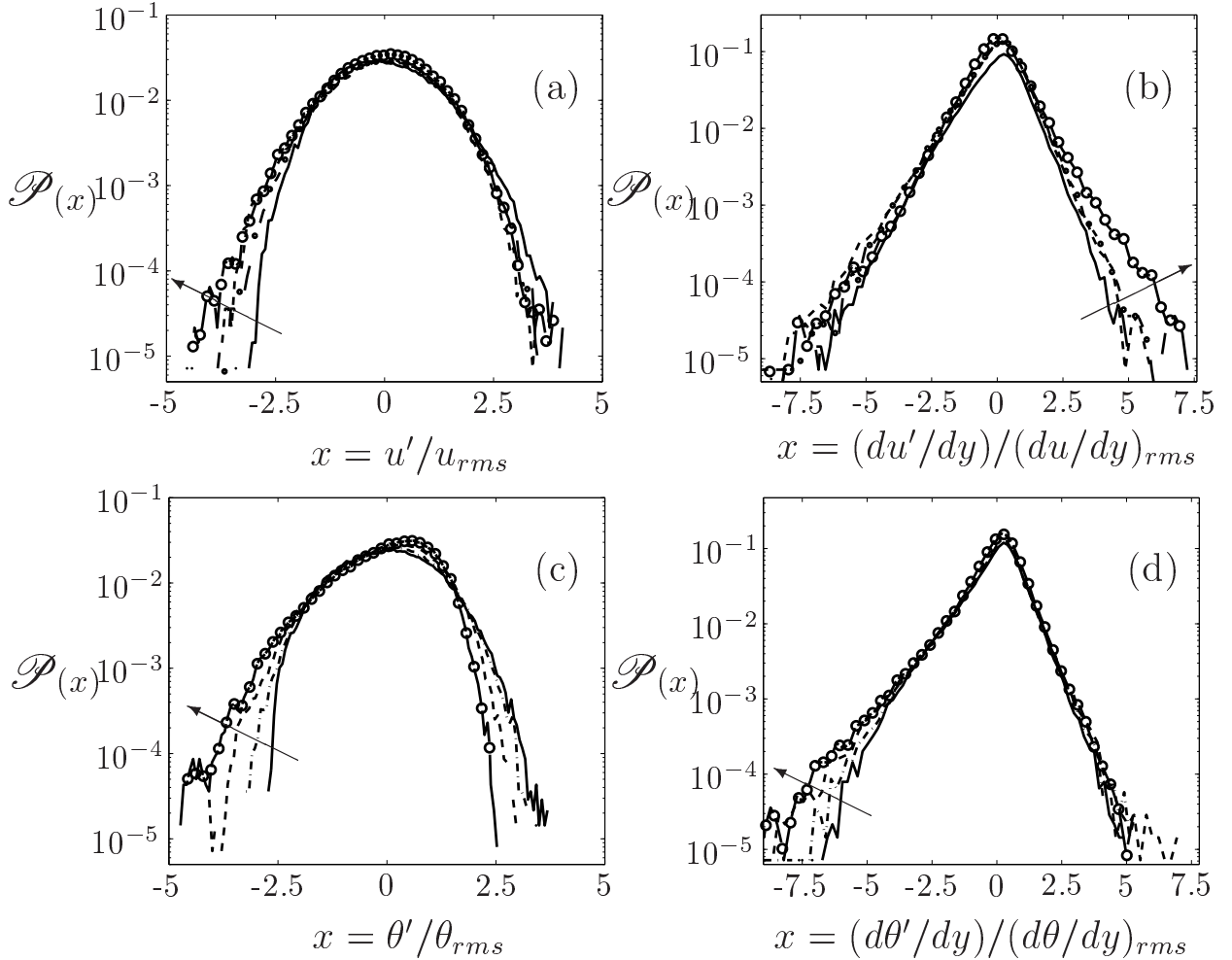}
\caption{
Probability density functions of (a) the streamwise velocity, (b) the wall-normal gradient of the streamwise velocity,
(c) the passive scalar, and (d) the wall-normal gradient of the passive scalar for the isothermal case I at the downstream
position $x/h = 25$. Unfilled circles, $y/h = 0.65$; dashed curves, $y/h = 1.3$; dotted-dashed curves, $y/h = 2.0$; and solid
curve, $y/h = 2.65$. Arrows point in the direction of decreasing $y/h$.
%
%
}
\label{fig:pdfu}
\end{figure}
\subsection{Probability density functions of the velocity and scalar fields}
\begin{figure}[thb]
\centering
\includegraphics[width=\linewidth]{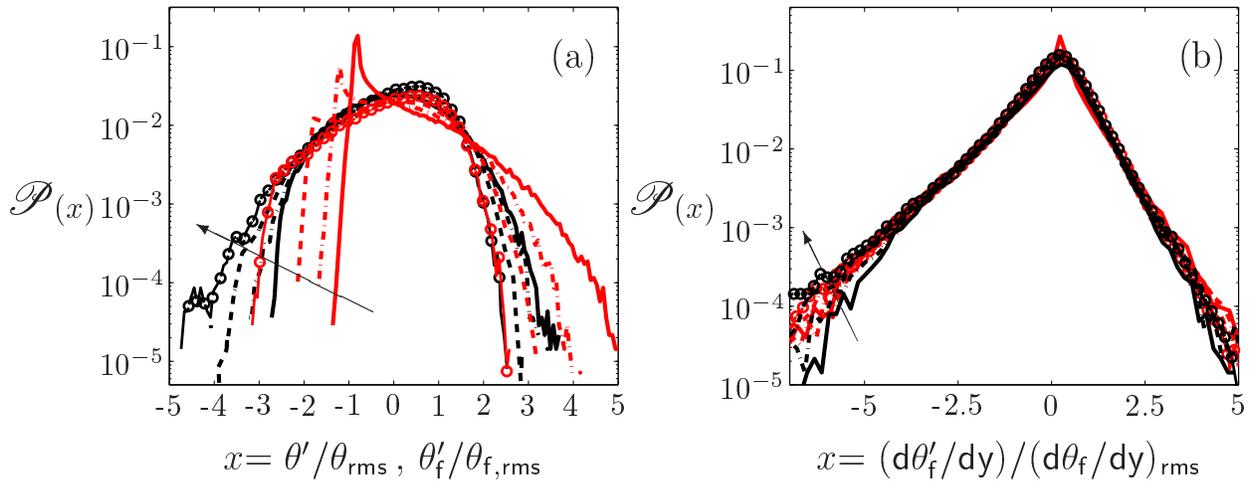}
\caption{
Probability density functions of (a) different species and (b) the wall-normal gradient of the scalars at $x/h = 25$
at several wall-normal distances; black color is used for the passive scalar and the red color for the fuel species. Unfilled
circles, $y/h = 0.65$; dashed curves, $y/h = 1.3$; dotted-dashed curves,$ y/h = 2.0$; solid curves, $y/h = 2.65$. Arrows point
in the direction of decreasing $y/h.$
}
\label{fig:pdfy}
\end{figure}

The PDFs of the streamwise velocity fluctuation and its wall-normal gradient are shown
in figures~\ref{fig:pdfu}(a) and (b), where a good collapse
of the PDFs in a large portion of the domain for both the velocity and its gradient is observed.
Similarly, the PDF of the passive scalar and its wall-normal gradient are shown
in figures~\ref{fig:pdfu}(c) and (d).
Nevertheless, rather different characteristics for the PDFs of the velocity and their gradients are
observed in the inner region, i.e. $y/y_{1/2} < 0.4$ or $y/h< 1.0$.
The PDFs of the gradients have longer tails in the near-wall region than those of the velocity.
Note that, when a quantity under consideration has an
intermittent nature, its PDF is very sensitive to the choice
of the range for the horizontal axis. 
For the PDFs of the velocity and scalar
derivatives in figure~\ref{fig:pdfu}, the horizontal range is chosen such
that outside this range, the scatter is very large and the probabilities are
lower than $10^{-5}$ indicating extremely rare events.
This makes the conclusions of the present investigation confined to this
particular range. Moreover, the PDF tails for the spatial derivatives are
known to depend on the resolution of the computation. In the present simulations,
as shown in figure~\ref{fig:kolm_tayl}(a), the resolution for the wall-normal distances
below $y/h =2.5$ compares well with the Kolmogorov scale and is better for the exothermic case.
For the wall-normal distances smaller than $y/h =1.5$ the grid spacing is equal or smaller
than the local Kolmogorov scale. The fine grid spacing is a clear and robust indication
that the PDFs are well represented and the results are thereby reliable.
It is interesting to compare the PDF of the velocity field to that of the passive scalar field.
At the extreme tails of the PDFs, a large difference between
the velocity gradients and scalar gradients are observed.
This is clear if we compare the absolute value of the PDFs' tail for large excursions from the mean.
These observations are better quantified by examining the
difference in the higher order moments at these particular positions. 
In Table~II, it is observed that the flatness
factors are substantially higher for the passive scalar compared to the velocity
components at the near-wall position, while the differences are small at $y_{1/2}$.
This is indeed consistent with the formation of much sharper gradients observed
in the scalar field, which is accentuated in the high-shear region near the wall.
The PDF of different species for the exothermic case is qualitatively similar to the isothermal case.
This is due to the fact that the chemical reaction is the dominating
effect regarding the difference between fuel and passive scalars.
The occurrence of a double peak close to the wall for the fuel
PDF and other characteristics are almost the same\cite{Pouransari2013},
which shows that the modes of the reacting scalars are
similar in the two cases with and without heat release.
Thus, for comparison of passive and reactive scalars, we may focus on
PDF shapes of only one simulation case, but at several wall-normal positions.
A comparison between the PDFs of the passive and reactive scalars for different wall-normal
positions, see figure~\ref{fig:pdfy}(a), shows that the PDFs of the two species are
very different at large scales and the chemical reaction plays an important role in the shape of the PDFs.
It can already be understood that the reaction effects on the species are much
larger than the heat release effects.

In figure~\ref{fig:pdfy}(a), the peaks of the fuel PDFs are clearly shifted towards
the lower concentrations and are sharper compared to those of the passive scalar.
It is observed that the PDFs for the two species hardly collapse at any particular wall-normal position.
Similar to the argument in relation to figure~\ref{fig:pdfu}, the
horizontal range in figure~\ref{fig:pdfy} is chosen such that outside
this range, the scatter is very high and events are extremely rare. 
The PDFs of the scalar gradients, the passive and reactive scalars, are barely
distinguishable throughout the entire wall-normal levels and a very good collapse
for all curves is obtained using the $rms$-values for normalizations, see figure~\ref{fig:pdfy}(b).
Thus, the turbulent mixing is overwhelming at the small scales and
the reaction does not play the key role at this level and universality of
the small-scales is recovered. This is confirmed further with the
good collapse achieved for the derivatives, see figure~\ref{fig:pdfy}(b).
This is an important observation made in the present study,
the fact that the PDFs of the gradients recover a fairly similar
shape for the passive and reactive scalars, is a clear indication of recovery of universality going
from large to small scales and is certainly true for the bulk of the gradient fluctuations.
However, a small deviation between the different cases can still be observed in 
the left tail of the PDFs in figure~\ref{fig:pdfy} (b).
This indicates a residual anisotropy (notice that for purely isotropic
statistics the PDF should be symmetric) and is connected to the fact that
the anisotropy might well be non-universal, depending on flow-specific features.
This important issue is addressed in the following part of the paper. 

\begin{figure}[!htbp]
\centering
\includegraphics[width=\linewidth]{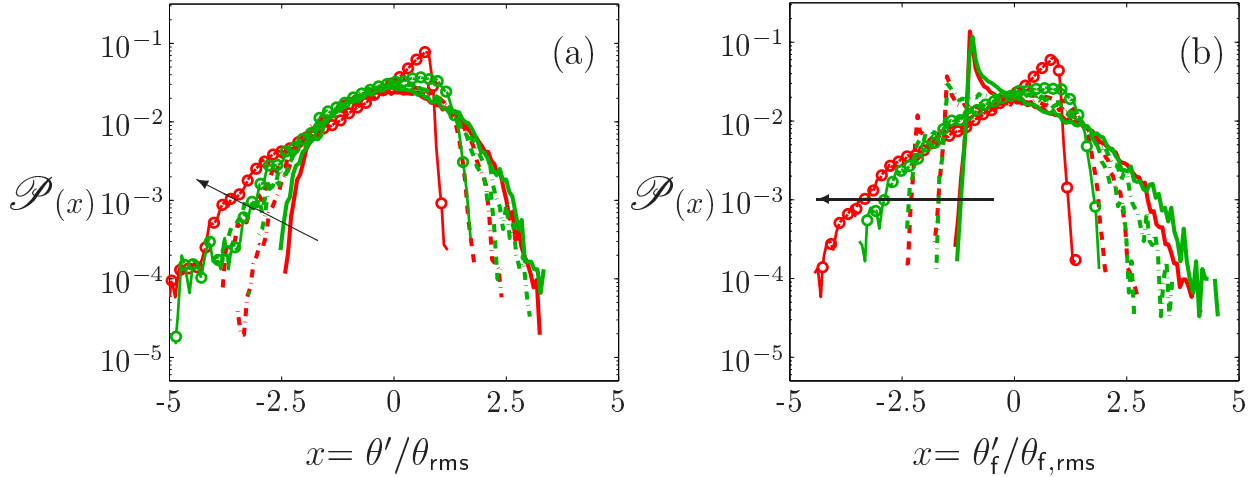}
\caption{\label{fig:pdf_Da} 
Probability density functions of (a) the passive scalar and (b) reactive scalars at $x/h = 25$ at several wall-normal
distances; red color is used for the higher Damkohler case II and the green color for the lower Damköhler case III; line styles
are as following: unfilled circles, $y/y_{1/2} = 0.25$; dashed curves, $y/y_{1/2} = 0.5$; dotted-dashed curves, $y/y_{1/2} = 0.75$; and
solid curves, $y/y_{1/2} = 1.0$. Arrows point in the direction of decreasing $y/y_{1/2}$.
}
\end{figure}
%
It is important to note that the reaction zone structure and the PDF of
the reactive scalars also depend on the Damk\"ohler number and
conclusions cannot be drawn without considering its influence.
Thus, to distinguish between the heat release and the Damk\"ohler number effects,
a comparison between two DNS cases with different Damk\"ohler numbers is performed.
Figure~\ref{fig:pdf_Da} shows the PDFs of both the passive and reactive scalars for these two cases.
The Damk\"ohler numbers are equal to $Da=500$ and $1100$ for cases~II and~III, see Table 1.
Four different wall-normal positions in terms of the local half-height of the jet $y_{1/2}$ are considered.
As expected, the turbulent flame becomes thinner with increasing Damk\"ohler number\cite{Pouransari2013},
however, we can expect a similar dynamical behavior for the flame at the
same wall-normal distance in terms of $y_{1/2}$. The two cases have similar characteristics at almost
the entire wall-normal direction, except in the near-wall region at $y_{1/2}=0.25$.
This is where the Damk\"ohler number effects are dominant and the PDF of case~II with
a higher Damk\"ohler number is more skewed toward the negative side and the
peak of PDF for high Damk\"ohler number case obtains a higher value.
This can also be associated with the isothermal wall cooling effects in the near-wall region.
\begin{figure}[!htbp]
\centering
\includegraphics[width=\linewidth]{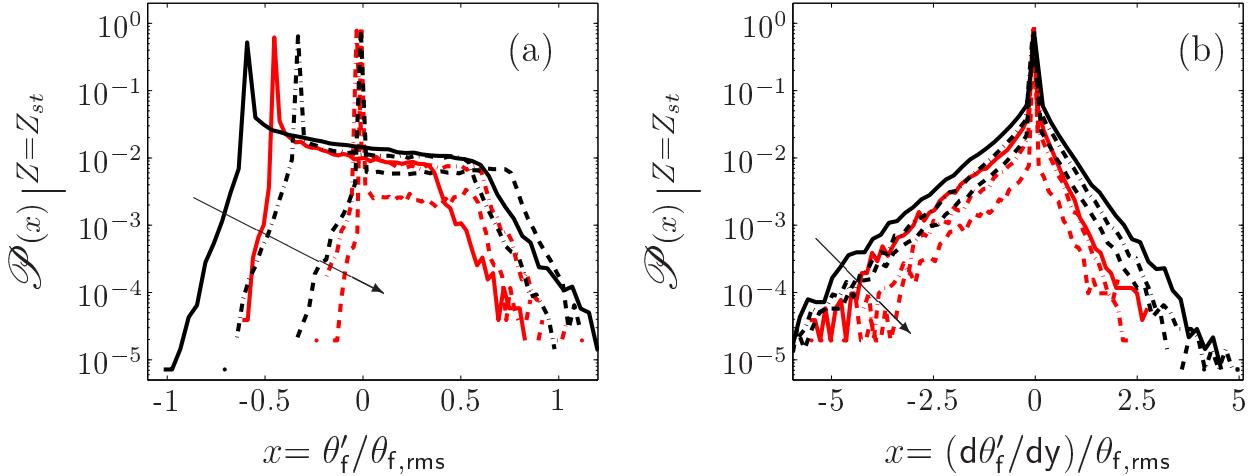}
\caption{\label{fig:pdf_cond} 
Probability density functions of (a) the reacting species and (b) the wall-normal gradient of the same scalars,
conditioned on the mixture fraction at $x/h = 25$ at several wall-normal locations; black color is used for the isothermal
case I and the red color for the exothermic case II; line styles are as following: dashed curves, $y/y_{1/2} = 0.5$; dotted-dashed
curves, $y/y_{1/2} = 0.75$; and solid curves, $y/y_{1/2}$ = 1.0. Arrows point in the direction of decreasing $y/y_{1/2}$, i.e., moving
closer to the wall.
}
\end{figure}

%
Furthermore, reactive scalars are better characterized using conditional statistics,
in particular those conditioned on the mixture fraction. The ramp-cliff contribution to the reactive scalar
statistics in the reaction zone is much larger than the rest of the mixture fraction field.
Thus, to better reflect the differences between the reactive scalars in the isothermal and
exothermic cases, the conditional PDFs on the mixture fraction are given in figure~\ref{fig:pdf_cond}. 
Three different wall-normal positions in terms of the local half-height of the jet are chosen for the two cases.
Note that the passive scalar concentration is used as the representative of the local mixture fraction.
The PDFs of the reactive scalars are conditioned on the local mixture fraction.
Around ten percent of the stoichiometric mixture fraction value $Z_{st}=1/(1+\phi\, r) =0.33$,
where $\phi=\theta_{F,inlet}/\theta_{O,inlet}$ is the equivalence ratio and $r =\nu_O/\nu_F =1$
as is used in the reaction equation. It is observed that if a narrower range is chosen for
the conditioning, then the PDF tails become more damped and the corresponding peaks
become sharper. It is notable that the PDFs of the reactive scalar derivatives,
as seen in figure~\ref{fig:pdf_cond}(b) are more skewed toward the negative side in
the conditional plots than the corresponding unconditional PDFs in figure~\ref{fig:pdfy} (b).
In particular, this is observed in the reaction zone and in the center of the
flame at $y/y_{1/2}=1.0$, which is similar to what is found in the anisotropic ramp-cliff like structures.
\begin{figure}[thb]
\centering
\includegraphics[width=\linewidth]{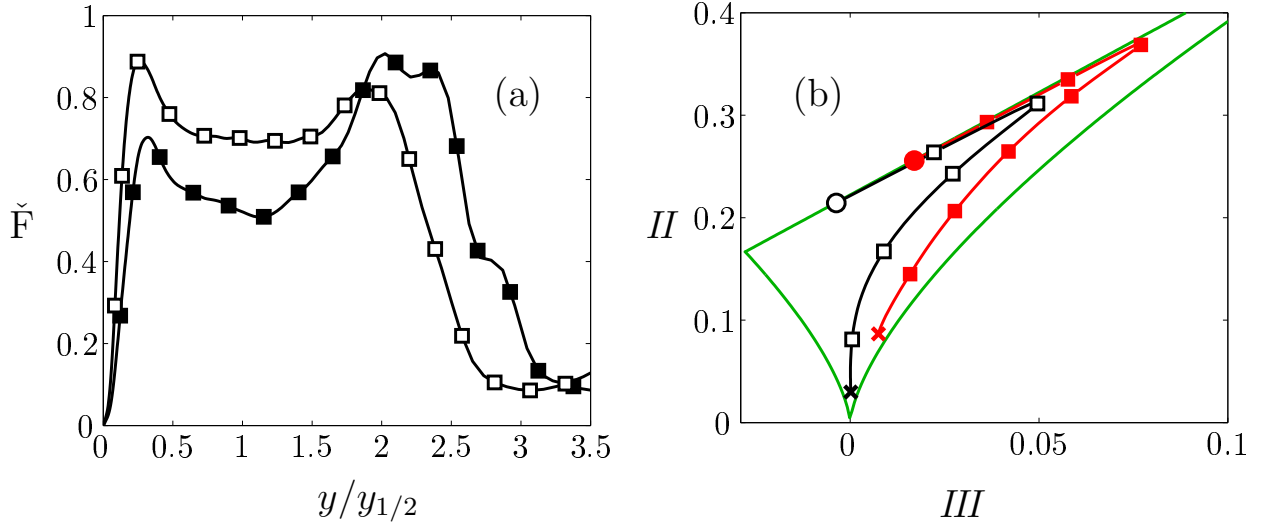}
\caption{
(a) Lumley flatness factor. Black unfilled squares, isothermal case I and black filled squares, exothermic case II. (b)
Anisotropy invariant maps at downstream position $x/h = 25 $for the two cases. The circle marks indicate values at the wall
and the cross signs show the corresponding values at the position of the maximum mean streamwise velocity for each case.
Black unfilled squares, isothermal case I and red filled squares, exothermic case II.
}
\label{fig:lumley_anisot}
\end{figure}
\section{Anisotropy invariant maps} \label{sec:anisot}
Before moving to the discussion on the higher order moments of the velocity and
scalar fields at large- and small-scales, let us have a closer look at a more standard
indicator of the anisotropic properties of the large-scale velocity components
through anisotropy of the second-order Reynolds stress tensor $b_{ij}$ defined as
\begin{equation}
b_{ij}= \frac{\oline{u_i''u_j''}}{2K}- \frac{1}{3}\,\delta_{ij},
\end{equation}
where $K$ is the turbulent kinetic energy and $\delta_{ij}$ is the Kronecker delta.
The second $I\!I$ and third $I\!I\!I$ invariants of $b_{ij}$ are defined as
\begin{equation}
I\!I=b_{ij}b_{ij}, \qquad I\!I\!I=b_{ik}b_{kj}b_{ji}.
\end{equation}
The Lumley flatness factor\cite{lumley1977,lumley1978} defined as
\v{F}$= 1+9\,I\!I+27\,I\!I\!I$ is used to study the turbulence characteristics. 
Its value varies between zero and one with small values indicating a
flow with characteristics of quasi two-dimensional turbulence and values close
to unity denote the structure of the isotropic three-component turbulence.
Figure~\ref{fig:lumley_anisot}(a) shows the value of \v{F}
for the two cases. As expected, \v{F} is small inside the boundary
layer for both cases and gradually increases and approaches unity away from the wall.
The maximum value \v{F}$_{max}=0.91$ for the isothermal case~I is higher than that for
the exothermic case~II where \v{F}$_{max}=0.69$.
This indicates that in both cases a quasi two-dimensional characteristic
is observed close to the wall for $y^+<9$. Away from the wall,
around $y/y_{1/2}=0.25$, the heat release due to the chemical
reaction tends to increase the flow anisotropy.
Hence gives a smaller peak for the exothermic case~II.
Further away from the wall, a weak local minimum forms at about $y/y_{1/2}=1.0$ for the two cases.
The local minimum is smaller for the exothermic case~II, indicating that the high amount of the combustion
heat release, present in the center of the jet flame, has increased the quasi two-dimensional character of the flow.

To further characterize the turbulence behavior of the flow field, the anisotropy invariant
maps\cite{lumley1977,lumley1978} are presented by cross-plotting the second $I\!I$ and third $I\!I\!I$
invariants.
The limits in the anisotropy map correspond to the quasi two-dimensional turbulence state when
$I\!I=2/9+2I\!I\!I$
and axisymmetric turbulence state when $I\!I=3/2(4/3|I\!I\!I|)^{2/3}$.
These limits define the anisotropy invariant map triangle within
which, all the physically realizable turbulence states lie.
The edges of this triangle are shown with solid lines in figure~\ref{fig:lumley_anisot} (b).
In the anisotropy invariant map, the lower corner of the triangle corresponds
to isotropic turbulence with zero anisotropy and axisymmetric turbulence
corresponds to the left and right edges of the triangle.
At the left edge, one component of velocity
fluctuations is smaller than the other two, while at the right edge, one component
of velocity fluctuations dominates the other two. The upper edge
of the triangle corresponds to the quasi two-dimensional turbulence.

Figure~\ref{fig:lumley_anisot}(b) shows the anisotropy invariant map
for cases~I and~II  at the downstream position $x/h=25$.
Starting from the wall, the trajectories of the two cases lie somewhere on the upper
edge, shown with circle marks in figure~\ref{fig:lumley_anisot}(b) and
turbulence has a quasi two-dimensional characteristic, as is the case close to any solid boundary.
However, the degree of anisotropy (invariant II) is larger for the case~II with
heat release, i.e. signifying a state closer to the one-component limit.
The trajectories move along the upper edge before they reach the
maximum of the invariant $I\!I\!I$, very close to the wall at $y^+=9$
and then return to lower values of the invariants.
The position of the maximum for both cases is the same, but
again the anisotropy magnitude is larger for the exothermic case.
Further away from the wall, trajectories move toward the lower
corner of the triangle and reach there individually at about the same wall-normal level
of the position of the maximum streamwise velocity or the local
zero shear around $y/y_{1/2}=0.25$, shown with crosses in figure~\ref{fig:lumley_anisot}(b).
This is consistent with the fact that the lower
corner of the anisotropy invariant map represents an isotropic
energy distribution of the turbulence, and indeed the
flow exhibits the most isotropic characteristics at the zero shear location.
Moving even further away from the wall,
in the region $0.2 \sim 0.25 <y/y_{1/2}<2 \sim 2.5$,
both trajectories remain close to the lower
corner of the map, initially with positive values of the invariant $I\!I\!I$ and then
gradually move away form the lower corner towards the left edge.
The trajectories reach the left edge of the map at about $y/y_{1/2}=3 \sim 3.5$, not shown in the figure.
Away from this position, it is well outside the jet boundary and all the turbulence intensities tend to vanish.
\section{Higher order moments}\label{sec:ho} 
In order to better quantify the degree of anisotropy of the flow, it is useful 
to identify a set of observables, which highlight either large- or small-scale field
properties, which would exactly vanish for purely isotropic turbulence.
For the passive, passive-reactive and passive-active scalars,
it suffices to take any odd moment of either the single point scalar or its derivatives.

In order to deal with dimensionless quantities, a normalization with a suitable
power of the $rms$-value is used, defining the so-called generalized skewness of order $n$: 
\begin{equation}
S^{(n)}_{\varphi} = \frac{\oline{(\varphi'')^{(2n+1)}}}{{\oline{(\varphi'')^2}}^{(2n+1)/2}},
\label{eq:sk}
\end{equation}
where with $\varphi'' = \varphi - \widetilde{\varphi}$ we denote the
fluctuation of any scalar field around its Favre-averaged mean value.
For isotropic statistics, we would have $S^{(n)}_\varphi=0$
for any $n$ (see e.g. Monim \& Yaglom\citep{monin_yaglom}).
In the following, we will limit ourselves to address the first non-trivial case, $n=1$,
which is typically called the skewness of a field and we will drop the dependency on the index $(n)$.
Concerning a vector field, the skewness defined in terms of any of its components
would be zero in an isotropic field, i.e. $S_{u,v,w} = 0$ where $u,v,w$ are
the streamwise, wall-normal and spanwise components, respectively.
The situation is different if one considers the gradients. For vector fields, only the odd moments of 
the transverse gradients must vanish in an isotropic ensemble, while the longitudinal ones
can be (and indeed are) different from zero also in a purely rotational invariant case.
For instance, if we focus on the wall-normal derivatives, we must expect that the skewness of 
$du/dy$ and $dw/dy$ are genuine measurements of the degree of anisotropy, while $dv/dy$ is not, 
being affected by both isotropic and anisotropic fluctuations.

Concerning a way to characterize the intense non-Gaussian properties of the field, 
independently of its isotropic or anisotropic 
content, it is useful to introduce the generalized flatness factor of order $n$:
\begin{equation}
F^{(n)}_{\varphi} = \frac{\oline{(\varphi'')^{(2n+2)}}}{{\oline{(\varphi'')^2}}^{(2n+2)/2}},
\label{eq:fl}
\end{equation}
which is a proxy of the degree of departure from a purely Gaussian shape. For instance for the simplest case $n=1$,
we have $F^{(1)}=3$ for a Gaussian variable.
%
\begin{figure}
\centering
\includegraphics{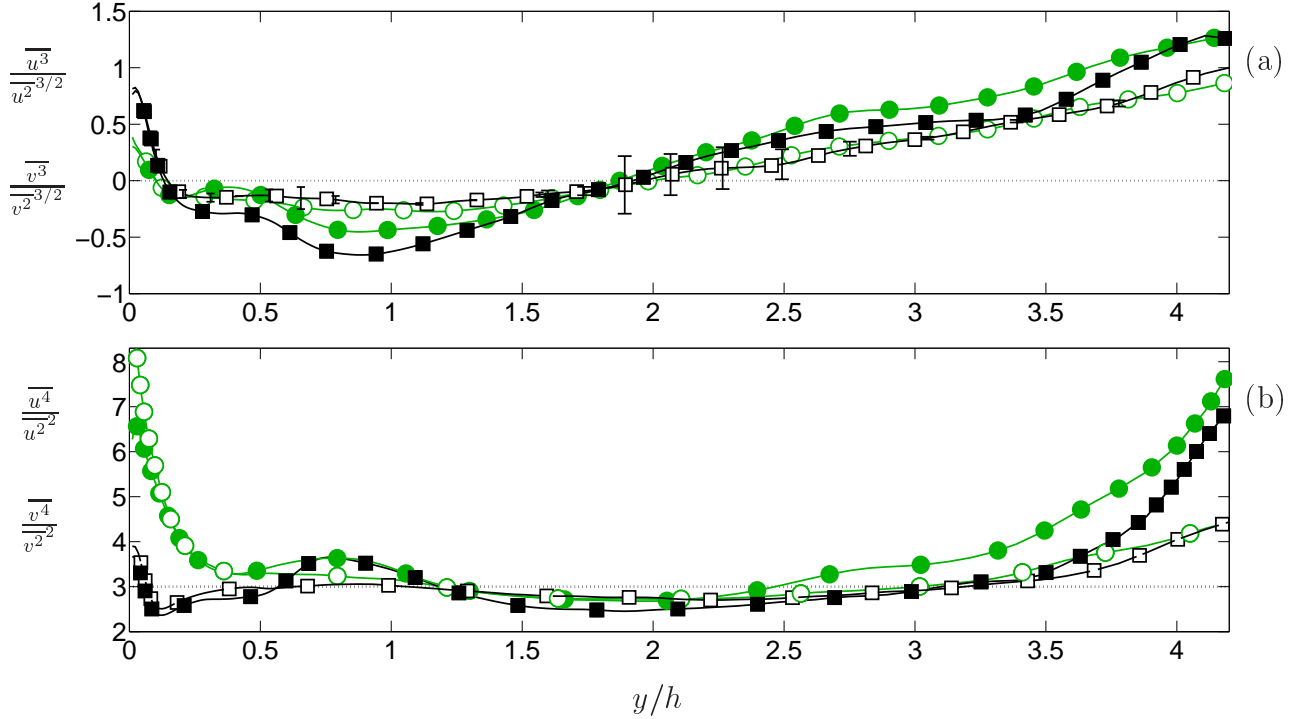}
\caption{
Distribution of third and fourth-order moments of the velocity fluctuations in the cross-stream direction at
downstream position $x/h = 25$; (a) skewness and (b) flatness. Black squares, streamwise velocity, 
$(\oline{u^3}/\oline{u^2})^{3/2}$, $(\oline{u^4}/\oline{u^2})^{2}$; green
circles, wall-normal velocity,
$(\oline{v^3}/\oline{v^2})^{3/2}$, $(\oline{v^4}/\oline{v^2})^{2}$.
 Unfilled and filled symbols denote the isothermal case I and exothermic case
II, respectively. In order to give the magnitude of error bars, we report the statistical oscillations only for a subset of points
for the streamwise component of skewness and flatness factors.
}
\label{fig:skfl_u}
\end{figure}

 In the sequel, we will limit ourselves to consider the case 
$n=1$ and we will therefore also in this case drop the dependency on $n$ in its symbol.
A fluctuating quantity with flatness larger than three will typically have a
PDF with {\it long flat tails}, with a probability to attain very intense fluctuations
that are higher than for a Gaussian distribution.
If the flatness factor becomes larger and larger with the change of a control parameter (e.g. the Reynolds 
number) the variable is said to be intermittent.

\begin{figure}
\centering
\includegraphics{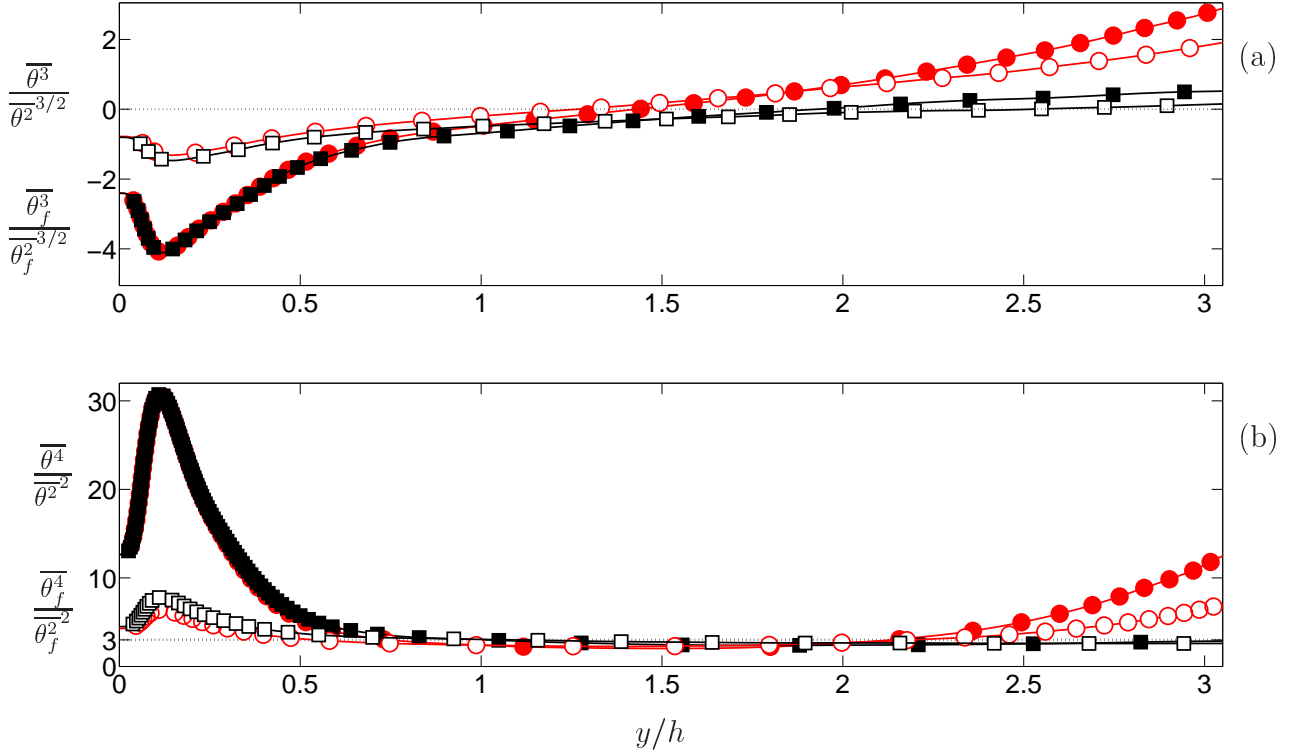}
\caption{
Distribution of the third and fourth-order moments of the scalar fluctuations in the cross-stream direction at
$x/h = 25$ for cases I and II; (a) skewness and (b) flatness. Black squares, passive scalar, $(\oline{\theta^3}/\oline{\theta^2})^{3/2}$, and
 $(\oline{\theta^4}/\oline{\theta^2})^{2}$; red circles, fuel, $(\oline{\theta_f^3}/\oline{\theta_f^2})^{3/2}$, and
 $(\oline{\theta_f^4}/\oline{\theta_f^2})^{2}$. Unfilled and filled symbols denote the isothermal case I and exothermic case II, respectively.
}
\label{fig:skfl_y}
\end{figure}
In order to investigate the intermittency and the anisotropies at different wall-normal positions, 
the skewness and the flatness
of the scalar concentration fluctuations, of the velocity components and
of their gradient fields are analyzed for the two different sets of simulations with and without heat release.
Note that, the skewness of the velocity or scalar fluctuations are connected to the
large-scale anisotropies and the skewness of the transverse gradients of the
velocity or scalar are linked to the small-scale anisotropies of the field.
\begin{table}
\begin{center}
\caption{The skewness and flatness factors at two particular wall-normal positions,
deduced from figures~\ref{fig:skfl_u},~\ref{fig:skfl_du},~\ref{fig:skfl_y} and \ref{fig:skfl_dy}.
The values are demonstrated at two locations, at $y^+ = 10$ i.e. $y/h = 0.11$,
which is a representative of the near-wall region and $y/y_{1/2} = 1.0$ i.e. $y/h =2.6$, which lies inside the peak of the jet flame.
The error bars are of the order of one unit in the first decimal, which means that the skewness 
of the spanwise component at $y/y_{1/2} = 1.0$ is compatible with zero.
The first value is for the isothermal case I and the second for the exothermic case II.}
\begin{tabular}{c c c c c c c c c c c c c}
\hline\hline
\\
&\quad\quad&$F(@ y^+ = 10)$ &\quad\quad& $F(@ y/y_{1/2} = 1.0)$ &\quad\quad& $S(@ y^+ = 10)$ &\quad\quad& $S(@ y/y_{1/2} = 1.0)$\\
\\
$u$ &\quad\quad& 2.51 / 2.37 &\quad& 2.78 / 2.70 &\quad& 0.16 / 0.10 &\quad& 0.20 / 0.41\\
$v$ &\quad\quad& 5.30 / 5.04 &\quad& 2.81 / 3.15 &\quad& -0.04 / -0.04&\quad& 0.25 / 0.53\\
$w$&\quad\quad& 3.34 / 3.34 &\quad& 2.89 / 3.11 &\quad& 0.06 / -0.06 &\quad& -0.04 / 0.07\\
\\
$du/dy$&\quad\quad& 3.41 / 3.19 &\quad& 5.21 / 5.55 &\quad& 0.46 / 0.33 &\quad& -0.68 / -0.98\\
$dv/dy$&\quad\quad& 4.45 / 4.45 &\quad& 4.12 / 4.19 &\quad& -0.27 / -0.37 &\quad& -0.39 / -0.39\\ 			
$dw/dy$&\quad\quad& 4.45 / 4.16 &\quad& 4.78 / 4.39 &\quad& 0.04 / -0.13 &\quad& 0.07 / -0.01\\				
\\
$\theta$ &\quad\quad& 7.79 / 30.7 &\quad& 2.60 / 2.60 &\quad& -0.142 / -4.10 &\quad& 0.02 / 0.33\\			
$\theta_f$&\quad\quad& 6.40 / 30.7&\quad& 4.06 / 5.80&\quad& -1.28 / -4.10 &\quad& 1.24 / 1.71 \\				 
\\
$d\theta/dy$ &\quad\quad& 9.79 / 25.7 &\quad& 5.84 / 6.33 &\quad& -1.62 / -3.74 &\quad& -1.00 / -1.39\\			
$d\theta_f/dy$&\quad\quad& 8.32 / 25.7 &\quad& 7.71 / 9.53 &\quad& -1.46 / -3.74 &\quad& 1.24 / -1.99\\
\\
\hline\hline				 
\\
\end{tabular} 
\end{center}
\label{tabl:fl_sk}
\end{table}
\subsection{Skewness and flatness factors of the velocity and scalar fields}
The skewness and flatness of the velocity fluctuations are shown 
in figures~\ref{fig:skfl_u}(a) and (b) as a function of the wall-normal 
direction at a fixed downstream position $x/h=25$.
The positive skewness and high flatness of the streamwise velocity close to the wall are indications of
anisotropic large-scale velocities and a sign of the high intermittency in this region.
This is a rather universal behavior of all wall-bounded flows, such as in turbulent channel 
flows\citep{arne,johansson2013,jacob2004}.

The zero crossings of the streamwise velocity skewness, as shown in figure~\ref{fig:skfl_u}(a),
coincide with the two maxima of the streamwise turbulence intensity,
see figure~\ref{fig:umax}(c), that indicates these positions for the isothermal case~I.
The location of the inner maximum at $y^+ \approx14$ is similar to that in turbulent boundary layers.
In the turbulent wall-jet flow, a second zero crossing occurs at $y/y_{1/2} \approx 0.75$ or $y/h \approx 2$,
where also the outer maximum of the streamwise turbulence intensity is located.
The occurrence of the zero crossing of the skewness factor and the local minimum in the flatness profiles, 
see figure~\ref{fig:skfl_u}(b), is an indication
that around this region, where turbulence intensity has a local maximum, the PDF shape is most similar to a Gaussian. 
The skewness of the wall-normal component of the velocity follows a trend similar to that for the streamwise component.
However, the magnitudes of the skewness factors are even more affected by the heat release effects
 for the wall-normal component. 
Error bars are determined in terms of statistical uncertainty, i.e. by dividing the whole temporal evolution in two subsets and comparing the variations between the two measurements. As a result, error bars become smaller for those regions, where turbulent fluctuations are small.
Error bars are shown only for the streamwise component of the isothermal case~I, which reflects the statistical uncertainty
of the datasets and the limitations of the present results for the other components as well.

As seen in figure~\ref{fig:skfl_u}(b), the wall normal velocity in the
near-wall region has a higher flatness factor than the other
two components and shows a clear similarity with other wall-bounded flows.
In the outer region of the wall-jet, large positive values of the flatness factors
are observed for all three components. This is a clear way to quantify the
intermittency at the interface between turbulent and laminar regions,
just as is the case at the edge of the turbulent boundary layer.
The effect of heat release is enlarging the flatness factors and enhancing
the intermittency of the velocity field.
The occurrence of large flatness factors around $y/h \sim 3$ and beyond this 
position indicates high intermittency even for large-scales of the velocity field.
It implies that a large number of realizations are needed to have a proper 
statistical convergence in this region.
Beyond $y/h\approx 3$ the higher-order moment curves have a non-smooth
character, which is associated with the statistical inaccuracies pertaining to the
fact that this region falls outside the jet flame and the statistics are quantitatively
very small, see figures~\ref{fig:umax} and~\ref{fig:ymean}.
However, as shown in figure~\ref{fig:skfl_u}, for the streamwise component of
the velocity the magnitude of the error bars are relatively small and for other components
the size of error bars are comparable (not shown for clarity of the figure).
Comparing the flatness factors of different velocity components among each other shows that the 
flow field is fairly isotropic
in the isothermal case. The differences become larger for the exothermic case. 
The flatness factors of different velocity components for the two cases are similar in
the near-wall region, but further out in the jet, the heat release effects start to show some influences.
\begin{figure}[!tbh]
\centering
\includegraphics[width=\linewidth]{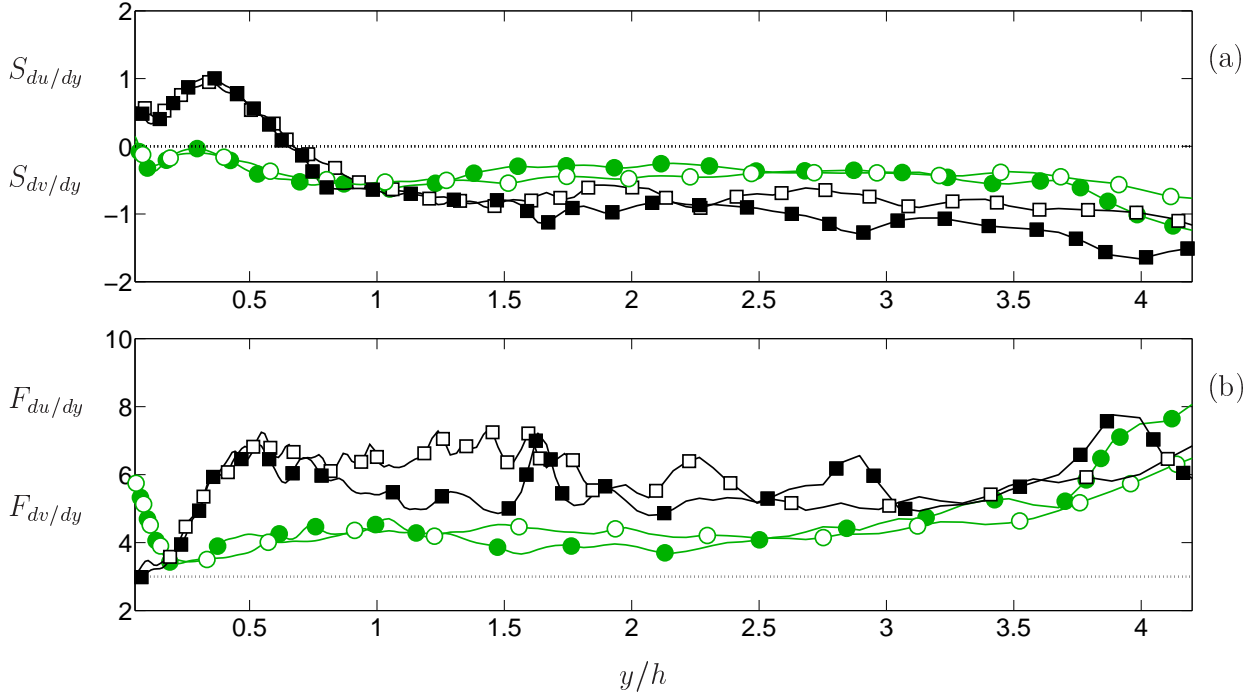}
\caption{
Distribution of skewness (a) and flatness (b) factors of the wall-normal gradient of the velocity fluctuation fields at
$x/h = 25$. Black squares, streamwise velocity, $S_{du/dy}$, $F_{du/dy}$; green circles, isothermal wall-normal velocity, $S_{dv/dy}$,
$F_{dv/dy}$. Unfilled and filled symbols denote the isothermal case I and exothermic case II, respectively.
%
}
\label{fig:skfl_du}
\end{figure}
\begin{figure}[!tbh]
\centering
\includegraphics[width=\linewidth]{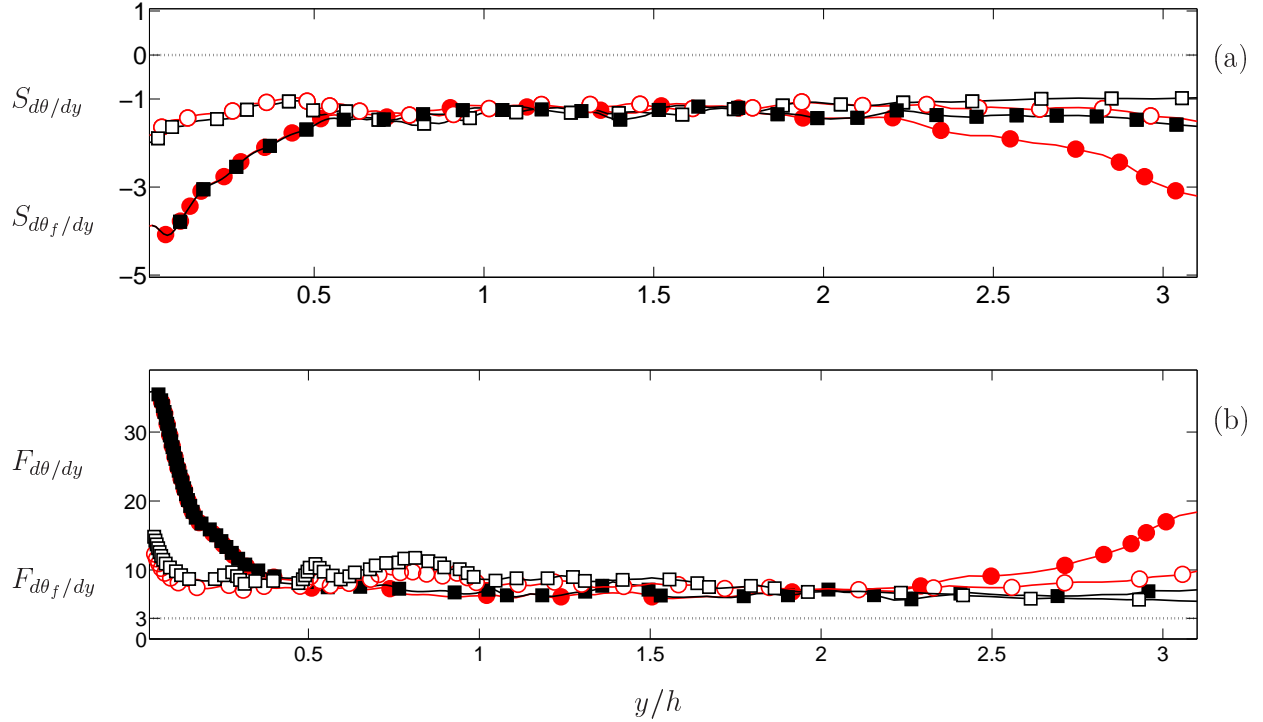}
\caption{
Distribution of (a) skewness and (b) flatness factors of the wall-normal gradient of the scalar fields at $x/h = 25$.
Black squares, passive scalar, $S_{d\theta/dy}$ and $F_{d\theta/dy}$; red circles, fuel, $S_{d\theta_f /dy}$ and $F_{d\theta_f /dy}$. Unfilled and filled symbols
denote the isothermal case I and exothermic case II, respectively.
%
}
\label{fig:skfl_dy}
\end{figure}

The skewness and flatness of the passive and reactive scalar fluctuations are shown in figure~\ref{fig:skfl_y}.
In the near-wall region and also within the jet center $0 < y/h < 2.5$, the 
heat release effects substantially dominate and the skewness factors for the two
scalars involved in exothermic case are much larger in magnitude than those in isothermal case.
On the other hand, the comparison between the skewness factors of the passive and reactive scalars 
for both the isothermal and exothermic cases, shows that the differences become significant away from the 
wall and also out of the jet center beyond $y/h > 2.5$. 
This implies that the effects of fuel consumption on the scalar 
skewness factors caused by the chemical reaction are comparable in the two cases.
The heat release leads to an increase of the largest scales of the flow in the 
wall-normal direction. Therefore, the effective mixing length of the flow is also increased. This results in a 
higher transport of fluid elements over a wider range in the cross-stream direction for exothermic
case compared to isothermal one.
Thus, as seen in figure~\ref{fig:umax}(c), larger fluctuation intensities form and the skewness values 
are also significantly larger, as seen in figure~\ref{fig:skfl_y}.
Close to the wall, the flatness factor of the passive and reacting scalars are comparable,
 but at the jet center, where most of the reaction takes place in the peak of the flame, $F(\theta_f)$ deviates 
significantly from $F(\theta)$.
This is indeed due to the fuel consumption, since at the peak of the
flame, the passive scalar does not experience any major changes, while fuel species are involved in the chemical 
reaction and face substantial gradients, and hence obtain a much larger flatness factor and intermittency.
For a more quantitative comparison between different components of the velocity and also different types of
 scalars in the two cases, the values of the skewness and flatness factors are summarized in Table~II. 
Two wall-normal positions are chosen, $y^+=10$, representative of the near-wall behavior 
and $y/y_{1/2} = 1.0$ that lies inside the jet flame.
\subsection{Skewness and flatness factors of velocity and scalar gradients}
%
The skewness and flatness factors of the velocity gradients in the wall-normal direction
are shown in figure~\ref{fig:skfl_du}, in order to illustrate characteristics dominated by small scales.
The skewness of the streamwise velocity gradient, $S_{du/dy}$ crosses zero close
to $y/y_{1/2} = 0.25$, where also the turbulent shear stress is zero, as seen in figure~\ref{fig:umax}(b).
For a wide region above this position at $0.4 < y/y_{1/2} < 1.4$ or $0.8 < y/h < 3.3$, there is a plateau with
a negative value of $S_{du/dy} \approx-0.5$ for the isothermal case and $S_{du/dy} \approx-1.0$
for the exothermic case, indicating a strong small-scale anisotropy.
Consistent with what was revealed by the anisotropy invariant maps,
heat release effects enhance the anisotropy of the flow field.
The anisotropy of the small scales seems to be of the same order as the anisotropy
of the large scales, but interesting enough, starting from the wall
region up to $y/y_{1/2} = 0.25$ or $y/h = 0.6$, $S_{du/dy}$ is positive, which indicates a
significant influence of the mean flow inhomogeneity down to the small scales. 
At the position, where the mean Reynolds shear stress (transversal shear) changes sign, $S_{du/dy}$
(transversal gradient) also changes sign.
The magnitudes of the skewness of the wall-normal and streamwise components of the
velocity gradients are comparable for the two cases, see rows four and five in Table~II. 
This indicates a quasi two-dimensional type of the turbulent flow near the wall.
All three components of the velocity gradients develop high flatness factors in 
the outer region, which is a further indication of
high non-Gaussian statistics at small-scales and coupled to the intermittency there.

The skewness and flatness factors of the scalar gradients in the wall-normal direction are 
shown in figure~\ref{fig:skfl_dy}.
A comparison between passive and reactive scalar gradients, last two rows in Table~II,
reveals that in the near-wall region, the effects of the combustion heat release are stronger than the chemical reaction itself.
Contrary, in the jet center, the differences between the two cases are small and the main difference is between the two scalars within the same simulation, meaning that the fuel consumption is the governing effect here.
It is remarkable that the anisotropy of the small scales is of the same order as the anisotropy of the large 
scales for the scalar field indicating a strong persistency of anisotropy if measured in this way. 
\section{Concluding remarks}\label{sec:conc}
We have addressed the problem of quantifying anisotropic and non-Gaussian fluctuations 
in a chemically reacting turbulent wall-jet flows at different combustion properties 
(isothermal and exothermic).
Fully passive, passive but reactive and active scalars are examined.
The chemical reaction and heat-release effects are found to be significant in different regions of the flow.
In the near wall-region, the heat release effects dominate, while in the
outer region of the jet, the chemical reaction influences are overwhelming.
By analyzing the statistics of the scalar fields, we have illustrated the importance of
combustion in the region far from the wall, where a very high intermittency is 
measured due to the depletion of the fuel.
The statistics of the gradients have been used to highlight the presence of a strong 'persistency of anisotropy'
at small-scales affecting both the velocity and scalars even at large distances from the wall.
The PDFs of the scalar gradients possess a rather similar shape for passive and
reactive scalars, indicating the recovery of universality for small-scale statistics.
Intermittent non-Gaussian fluctuations are found to be strong also for isotropic components,
with flatness for scalar gradients reaching values
around 10 in the bulk (and larger close to the wall). 
The Damk\"ohler number effects are observed to be influential in the near-wall region.
The conditional statistics suggest the influence of ramp-cliff like structures on the anisotropies.
Moreover, comparing the anisotropy invariant maps for the isothermal and exothermic
shows a clear effect of the heat release throughout the entire wall-normal direction.
Besides, the anisotropy levels for the large and small scales are compared using
the higher order moments and it is observed that the
heat release effects at the small-scale anisotropy are less pronounced.
The combined effects of strong intermittent features and strong persistency of
anisotropy at small scales have implications for the development of more accurate subgrid-scale models.

\begin{acknowledgments}
{\small{
Computer time provided by the Swedish National Infrastructure for 
Computing (SNIC) and the financial support 
from the Swedish National Research program of the Centre for Combustion 
Science and Technology (CECOST) are gratefully acknowledged.
L.B. acknowledges partial funding from the European Research Council under
the European Community's Seventh Framework Program, ERC Grant Agreement No. 339032.
The authors would like to thank the anonymous reviewers for their constructive
comments and suggestions.
}}
\end{acknowledgments}
\providecommand{\noopsort}[1]{}\providecommand{\singleletter}[1]{#1}%
%
\end{document}